# Robust Lineage Reconstruction from High-Dimensional Single-Cell Data

Gregory Giecold[1,2], Eugenio Marco[1,2], Lorenzo Trippa[1,2] and Guo-Cheng Yuan[1,2,3,*]

[1] Department of Biostatistics and Computational Biology, Dana-Farber Cancer Institute, Boston 02215, MA, USA

[2] Department of Biostatistics, Harvard T.H. Chan School of Public Health, Boston 02115, MA, USA

[3] Harvard Stem Cell Institute, Cambridge 02138, MA, USA

[*] To whom correspondence should be addressed. Tel: 1-617-582-8532; Fax: 1- 617-632-2444; Email: gcyuan@jimmy.harvard.edu

Present Address: Eugenio Marco, Editas Medicine, Cambridge 02142, MA, USA

## ABSTRACT

Single-cell gene expression data provide invaluable resources for systematic characterization of cellular hierarchy in multi-cellular organisms. However, cell lineage reconstruction is still often associated with significant uncertainty due to technological constraints. Such uncertainties have not been taken into account in current methods. We present ECLAIR, a novel computational method for the statistical inference of cell lineage relationships from single-cell gene expression data. ECLAIR uses an ensemble approach to improve the robustness of lineage predictions, and provides a quantitative estimate of the uncertainty of lineage branchings. We show that the application of ECLAIR to published datasets successfully reconstructs known lineage relationships and significantly improves the robustness of predictions. In conclusion, ECLAIR is a powerful bioinformatics tool for single-cell data analysis. It can be used for robust lineage reconstruction with quantitative estimate of prediction accuracy.

## INTRODUCTION

Over the past few years, high-throughput sequencing, flow and mass cytometry, microfluidics along with other technologies have evolved to the point that the measurements of gene expression are now possible at the single-cell resolution (1), providing an unprecedented opportunity to systematically characterize the cellular heterogeneity within a tissue or cell type. The high-resolution information of cell-type composition has also provided new insights into the cellular heterogeneity in cancer and other diseases (2). Single-cell data present new challenges for data analysis, and computational methods for addressing such challenges are still under-developed (3). Here we focus on a common challenge: to infer cell lineage relationships from single-cell gene expression data. While several methods have been developed (4-8), one common limitation is that the resulting lineage is often sensitive to various factors including measurement error, sample size, and the choice of pre-processing methods. However, such sensitivity has not been systematically evaluated.

Ensemble learning is an effective strategy for enhancing prediction accuracy and robustness that is widely used in science and engineering (9,10). The basic idea is to aggregate information from multiple prediction methods or subsamples. This approach has also been applied to unsupervised clustering, where multiple clustering methods are applied to a common dataset and consolidated into a single, more robust partition called the consensus clustering (11).

Here we apply such an ensemble strategy to aggregate information from multiple estimates of lineage trees. We call our method ECLAIR, which stands for Ensemble Cell Lineage Analysis with Improved Robustness. We show that ECLAIR improves the overall robustness of lineage estimates. Moreover, ECLAIR provides a quantitative evaluation of the uncertainty associated with each inferred lineage relationship, providing a guide for biological validation.

## MATERIAL AND METHODS

ECLAIR consists in three steps: 1. ensemble generation; 2. consensus clustering, and 3. tree combination. An overview of our method is shown in Figure 1.

**Ensemble generation**

Given a dataset, we generate an ensemble of partitions out of a population of $n$ cells by subsampling, which can be either uniform or non-uniform. For large sample size, we prefer to use a non-uniform, density-based subsampling strategy. Specifically, a local density at each cell is estimated as the number of cells falling within a neighbourhood of fixed size in the high-dimensional gene expression space. If the local density is above a threshold value, a cell is sampled with a probability that is inversely proportional to the local density; otherwise, the cell is always included. The resulting subsample exhibits a nearly uniform coverage of the gene expression space while retaining outliers in the cell population.

Each subsample is divided into clusters with similar gene expression patterns. The specific clustering algorithm is determined by the user and can be selected from $k$-means (12), affinity propagation (13), or DBSCAN (14). In practice, we find that $k$-means clustering typically offers a good balance between robustness of the final estimates and computational costs.

For a given clustering solution, a fully connected graph is constructed by connecting every cluster pair, with the edge weight defined as the average Euclidean distance (in the gene expression vector space) between all pairs of cells straddling the two clusters. A minimum spanning tree (MST) is defined as the tree connecting all clusters with minimal total weight. We use Prim's algorithm (15) to identify the MST. Later on, the path length between a pair of cells, $x$ and $y$, is defined as the minimum number of edges along the a MST path connecting their corresponding clusters, and denoted by $L(x,y)$.

The above clustering and linkage procedure is repeated $M$ times, each corresponding to a random subsample. At each iteration the resulting clusters are expanded to incorporate every cell in the population: each cell that has not been subsampled is assigned to its closest cluster. In the end, each tree in the ensemble provides a specific estimate of the lineage tree for the entire cell population.

Our goals are to aggregate information from the ensemble and to obtain a robust estimate of the lineage tree.

**Consensus clustering**

We start by aggregating the clustering information, searching for a consensus clustering that is on average the most consistent with the different $M$ partitions in the ensemble, using a strategy proposed by Strehl and Ghosh (11). For a population of n cells, the similarity between a pair of clusterings, $\lambda^{(a)}$ and $\lambda^{(b)}$, which contains $k^{(a)}$ and $k^{(b)}$ clusters respectively, is quantified by the normalized mutual information (NMI), defined as:

$$\phi^{(\text{NMI})}(\lambda^{(a)}, \lambda^{(b)}) = \frac{\sum_{h=1}^{k^{(a)}} \sum_{\ell=1}^{k^{(b)}} n_{h,\ell} \log\left(\frac{n \cdot n_{h,\ell}}{n_h^{(a)} n_\ell^{(b)}}\right)}{\sqrt{\left(\sum_{h=1}^{k^{(a)}} n_h^{(a)} \log \frac{n_h^{(a)}}{n}\right)\left(\sum_{\ell=1}^{k^{(b)}} n_\ell^{(b)} \log \frac{n_\ell^{(b)}}{n}\right)}}.$$

Here $n_h^{(a)}$ and $n_l^{(b)}$ denote the numbers of cells in the corresponding clusters, and $n_{h,l}$ stands for the number of cells in their intersection.

For an ensemble of $M$ partitions, $\lambda^{(1)}, \cdots, \lambda^{(M)}$, the consensus clustering $\lambda^¿ = \{C_1^¿, \cdots, C_K^¿\}$ is defined as the one that maximizes the average NMI with the $M$ partitions in the ensemble. The solution is computed by combining three approximation algorithms, CSPA, HGPA and MCLA, and selecting the one that performs the best (11).

**Tree combination**

The final step of ECLAIR amounts to constructing a representative tree connecting the consensus clusters. We first construct a fully connected graph $G^¿$, with the weight of the edge connecting clusters $C_i^¿$ and $C_j^¿$ given by

$$W_{ij} = \frac{1}{n_i^¿ n_j^¿ M} \sum_{x \in C_i^¿} \sum_{y \in C_j^¿} \sum_{m=1}^{M} L^{(m)}(x, y)$$

where $n_i^¿$ and $n_j^¿$ refer to the number of cells in $C_i^¿$ and $C_j^¿$, respectively, and $L^{(m)}(x, y)$ is the path length between cells $x$ and $y$ in the $m$-th tree. From this graph, again, we extract a MST, $T^¿$, by using Prim's algorithm. This is from now on referred to as the ECLAIR tree. In the main text, we show that the ECLAIR tree provides a robust estimate of the lineage relationship.

**Tree visualization**

We use the igraph Python package (http://igraph.org/python/) to visualize the various trees generated by SPADE or ECLAIR. To facilitate visualization, we encode the overall gene expression pattern associated with a cell cluster in a particular coloring scheme. Specifically, the raw gene expression pattern is subjected to Principal Component Analysis (PCA). The first three components are rescaled to the [0, 1] interval, and together define a unique color in the RGB encoding scheme. As such, clusters with similar expression patterns will have similar colors. Besides, the size of each node is scaled according to the number of cells it contains.

**Lineage tree comparison and robustness estimation**

When we compare two lineage trees, we need to compare not only their edge connections but also their node (i.e., cell cluster) identities, since the variation associated with subsampling results in different cell clusters. Although there exists a body of literature on graph comparison (16), we are not aware of any method that takes into account the node differences. We have therefore developed new metrics that are suitable for comparing lineage trees.

First, we define a metric to compare the overall similarity between two lineage trees: $T_1$ and $T_2$. For each tree, we evaluate the path length between every pair of cells in the population, based on the edge connectivity. The correlation between the two sets of path length values is used as a metric to compare the overall similarity of $T_1$ and $T_2$.

Second, we define edge-specific dispersion rates to evaluate the robustness of each edge within a lineage tree $T^i$. Specifically, for each edge $E_{ij}$ connecting a pair of clusters $C_i^i$ and $C_j^i$, we define the dispersion rate $D_{ij}$ associated with $E_{ij}$ as the standard deviation of path length $L^{(a)}(x,y)$, where $x$ and $y$ are randomly selected from $C_i^i$ and $C_j^i$ respectively, and $T^{(a)}$ is randomly selected from the ensemble of trees. To estimate the dispersion rate $D_{ij}$, we use the standard bootstrap procedure, creating 50 ECLAIR trees each obtained from a different initial subsample. The edges associated with lower $D_{ij}$ values are more reproducible and therefore more likely to reflect true lineage relationship.

Similarly, we define an intra-ensemble dispersion rate $\widehat{D}_{ij}$ to quantify the variation within the tree ensemble, as the standard deviation of path length across the individual trees within the ensemble. To distinguish the two terms, we call $D_{ij}$ the inter-ensemble dispersion rate. As shown in the main test, the values $\widehat{D}_{ij}$ and $D_{ij}$ are highly correlated. Since $\widehat{D}_{ij}$ can be evaluated much more efficiently compared to $D_{ij}$, it can be used as a quick and simple approximation.

**Numerical implementation and software package**

ECLAIR is implemented as an open-source Python package, which can be accessed at https://www.github.com/GGiecold/ECLAIR. In order to manipulate large datasets, we have made a number of efforts to optimize storage and numerical efficiency, including: a scalable Python module utilizing the HDF5 data structure, a sparse-matrix and streamlined implementation of the Strehl-Ghosh approximation algorithms for consensus clustering, along with scalable and efficient implementations of the hierarchical clustering, affinity propagation, and DBSCAN algorithms. All packages are accessible from either the aforementioned Github website or the Python Package Index (PyPI).

**RESULTS**

**Reconstruction of cell lineages in mouse embryos**

We used ECLAIR to analyze a public qPCR dataset, which contains gene expression information for 48 genes in 438 cells isolated from early mouse embryos (17). Previously, we developed a method called SCUBA to reconstruct and experimentally validate the cell lineages based on the temporal information (8). Here we applied ECLAIR to reanalyze this dataset without using the temporal information. The clustering was done by using *k*-means, with *k* set to 11 as suggested by a gap statistics analysis (18). We generated 50 trees in total, each obtained from a subsample containing 75% of the total number of cells. As shown in Figure 2, the SCUBA and ECLAIR trees have strikingly similar overall structures, and the corresponding nodes have similar gene expression patterns, which speaks of the high accuracy of the ECLAIR algorithm.

**ECLAIR significantly improves robustness over SPADE**

To evaluate the robustness of ECLAIR, we analyzed a publicly-available mass cytometry dataset (19), which contains the expression levels of 9 protein markers for 500,000 cells from the mouse hematopoietic system. We compared our results with those spawn by SPADE (4), one of the prevalent methods for lineage reconstruction. Despite its wide applications, it has been noted that two independent runs of SPADE often lead to significantly different outcomes. In the latest version, the developers attempted to solve the problem by fixing the value of a random seed for subsampling. However, this simple approach does not resolve the problem of intrinsic variability associated with respective data sampling.

We compared the performance of ECLAIR and SPADE based on the correlation between pairwise path lengths. For each method, we generated 50 lineage trees. For fair comparison, we set the down-sampling parameter to 50% for both SPADE and ECLAIR. For SPADE, each subsample resulted in one lineage tree; whereas for ECLAIR, a lineage tree was aggregated from 50 individual trees, each obtained from one subsample. To quantify the reproducibility of each method, we compared the cell-pair path lengths obtained from two different trees (Figure 3) (see Material and Methods for details). It is clear that the distribution is more densely populated near the diagonal for ECLAIR (Figure 3A) compared to SPADE (Figure 3C). For ECLAIR, the correlation coefficient varies between 0.73 and 0.94, with a mean of 0.86 and standard deviation of 0.05. For SPADE, the correlation coefficient varies from 0.70 and 0.83, with a mean of 0.75 and a standard deviation of 0.03, indicating ECLAIR significantly improves reproducibility.

In order to evaluate the robustness with respect to training data differences, we randomly divided the whole cell population into three equal-size non-overlapping subsets, labelled S1, S2, and S3, respectively. For each method, we constructed two lineage trees: one using S1 as the training set, whereas the other using S2 for model training. S3 was reserved as the testing set. Again, we compared the cell-pair path lengths obtained from two different trees from each method, based on cells in the S3 subset. The distribution is more densely populated near the diagonal for ECLAIR (Figure 3B) compared to SPADE (Figure 3D). This procedure was repeated 10 times. For ECLAIR, the correlation between different training sets varies from 0.72 to 0.91, with a mean value of 0.82; whereas for SPADE, the correlation varies between 0.62 and 0.82 with an average value of 0.75. Again, ECLAIR is significantly more reproducible compared to SPADE.

**Edge-specific dispersion rate**

Until now we have represented lineage as a binary relationship. However, this representation does not provide information about the uncertainty associated with lineage inference. In view of the significant variation across the lineage trees obtained from different methods or, as we have shown here, from different training sets even using the same method, it is important to systematically quantify the robustness of different edges and to identify those edges that are the most robust. To this end, we have developed a quantitative metric, called dispersion rate, to evaluate the robustness of each edge in a lineage tree (see Material and Methods for detail).

We then evaluated the edge-specific dispersion rates for the ECLAIR trees generated for the mass cytometry dataset. As described in the previous section, we obtained 50 ECLAIR trees, each from a randomly selected subsample containing 50% of the cells. To estimate the dispersion rates for each tree in the ensemble, we estimated the standard deviation of the path lengths based on the other 49 trees. Comparing the structure of two randomly picked ECLAIR trees (Figure 4A and Figure 4B), we find that the edges associated with lower dispersion rates (thicker edges) are conserved between the trees, indicating that the dispersion rate is an informative metric for edge robustness. For comparison, we also show two randomly picked SPADE trees obtained from the same data (Figure 4C and Figure 4D). We see that the overall structure is more variable.

Since each ECLAIR tree is derived from an ensemble of individual trees, the variability within the ensemble is also an indicator of edge robustness. To compare the intra-ensemble variability and the inter-tree variability, we defined an intra-ensemble dispersion rate for each edge (see Material and Methods for detail). We compared the intra-ensemble dispersion and dispersion rate for the ECLAIR tree (Figures 5, S1). The correlation between the intra-ensemble and inter-ensemble dispersion rates is 0.69, indicating that the intra-ensemble variation indeed can be used a good predictor for edge-specific robustness.

**DISCUSSION**

One important goal in single-cell analysis is to map the cellular hierarchy within a cell population. Computational methods have played an important role in inferring lineage relationships, but it must be reckoned that such predictions are often associated with a high degree of uncertainty. Our ECLAIR method provides a systematic way to evaluate the uncertainty associated with lineage reconstruction. By comparing with SPADE, a state-of-the-art method for lineage reconstruction, we show that our method has improved the overall robustness and further quantifies the uncertainty for each predicted lineage relationship. The most reliable link may be prioritized for further experimental validation.

Our ECLAIR analysis still does not entirely remove the variation even when using the same training dataset. However, if the ensemble size were set to a much large size, it follows from Cramer's theorem and techniques from large deviation theory that the ECLAIR tree converges to a single solution, although the rate of convergence is slow and at odds with most practical purposes (20).

To aid with biological interpretation, we have represented the lineage relationship as a tree, as is commonly done. On the other hand, the full graph contains additional useful information, and it is more reproducible than the minimum spanning tree estimate itself (Figure S2). The average correlation of graph weights is 0.96, which is much higher than that for the ECLAIR tree path lengths ($R = 0.86$). This is because the tree structure is sensitive to small perturbations of the underlying data (Figure S3).

Our ECLAIR method has important limitations. As in other similar methods, we assume that gene expression similarity is mainly contributed by lineage similarity. However, this ignores the contributions of other mechanisms such as spatial organizations. As such, we should only view the predicted cell lineages as one possible hypothesis. More comprehensive experimental data are required to more accurately define cell lineage information.

## ABBREVIATIONS

ECLAIR: <u>E</u>nsemble <u>C</u>ell <u>L</u>ineage <u>A</u>nalysis with <u>I</u>mproved <u>R</u>obustness; SPADE: Spanning-tree Progression Analysis of Density-normalized Events; MST: Minimum Spanning Tree; DBSCAN: Density-based Spatial Clustering of Applications with Noise; SCUBA: Single-cell Clustering Using Bifurcation Analysis; PCA: Principal Component Analysis; qPCR: quantitative Polymerase Chain Reaction; NMI: Normalized Mutual Information; RGB: Red-Green-Blue.

## FUNDING

This work was supported National Institutes of Health [grant number R01HL119099], and a Claudia Barr Award to G.C.Y. Funding for open access charge: National Institutes of Health.

## ACKNOWLEDGEMENT

We thank the members of the Yuan Lab for helpful discussions.

Figure 1

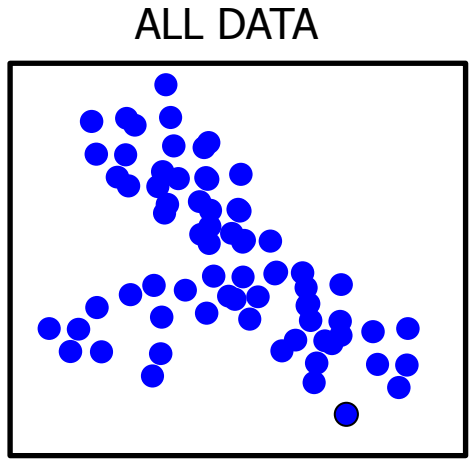
ALL DATA

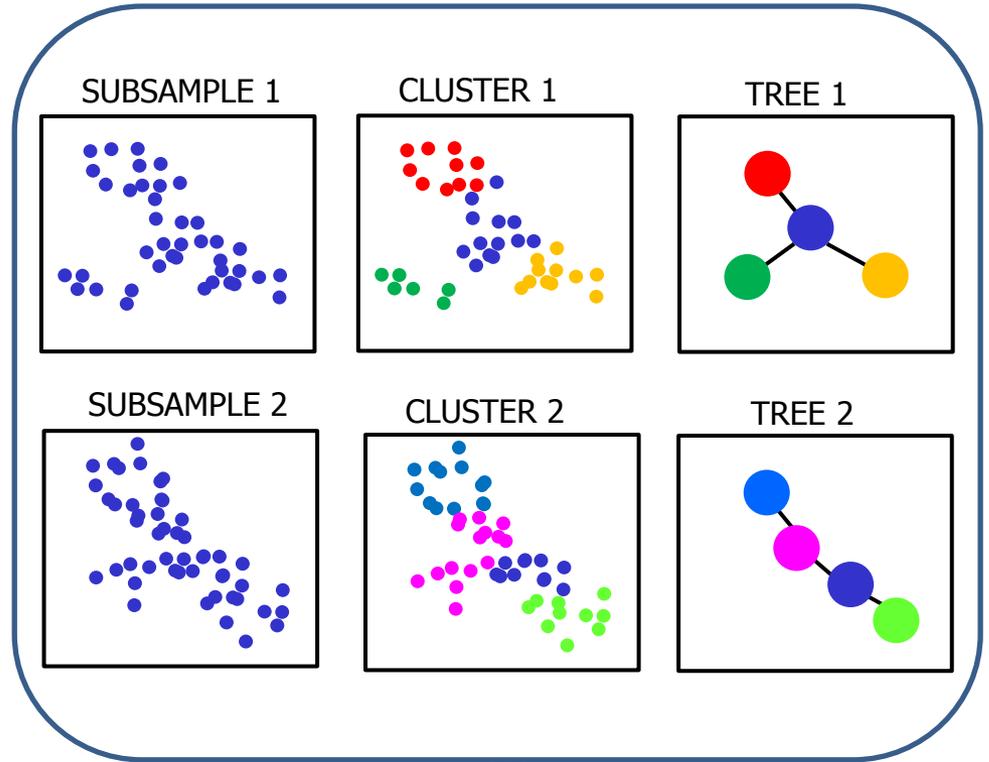

SUBSAMPLE 1 | CLUSTER 1 | TREE 1
SUBSAMPLE 2 | CLUSTER 2 | TREE 2

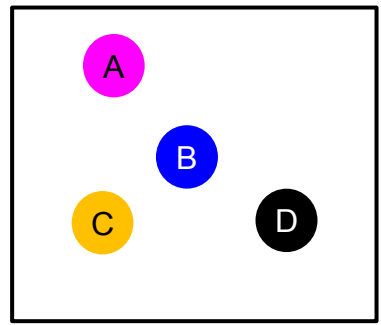
CONSENSUS CLUSTERING

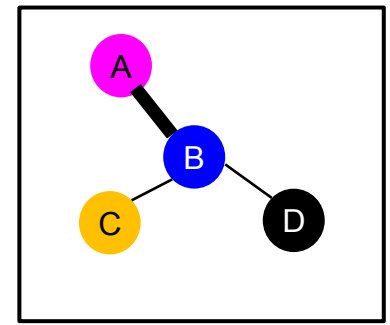
CONSENSUS TREE

Figure 2

A 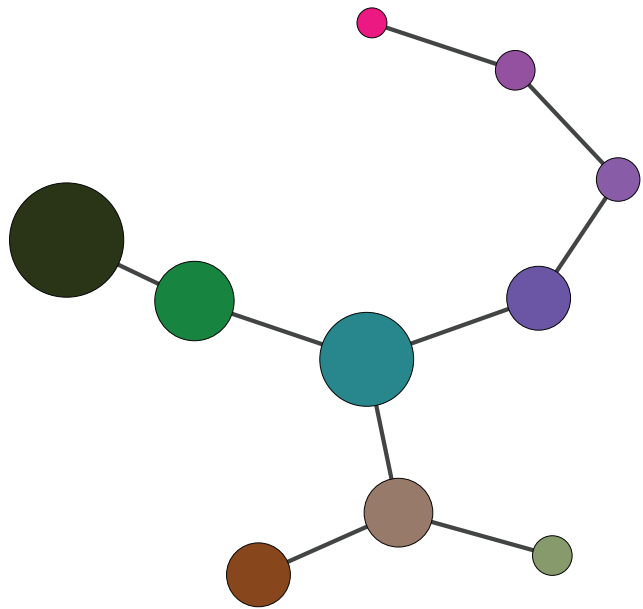

B 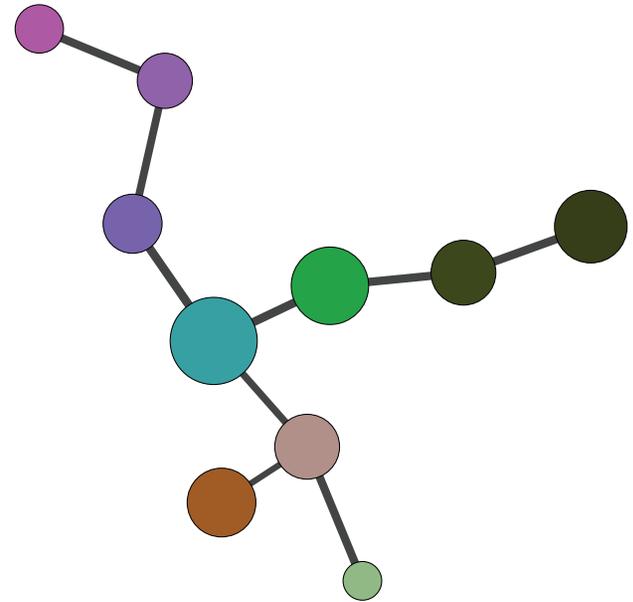

Figure 3

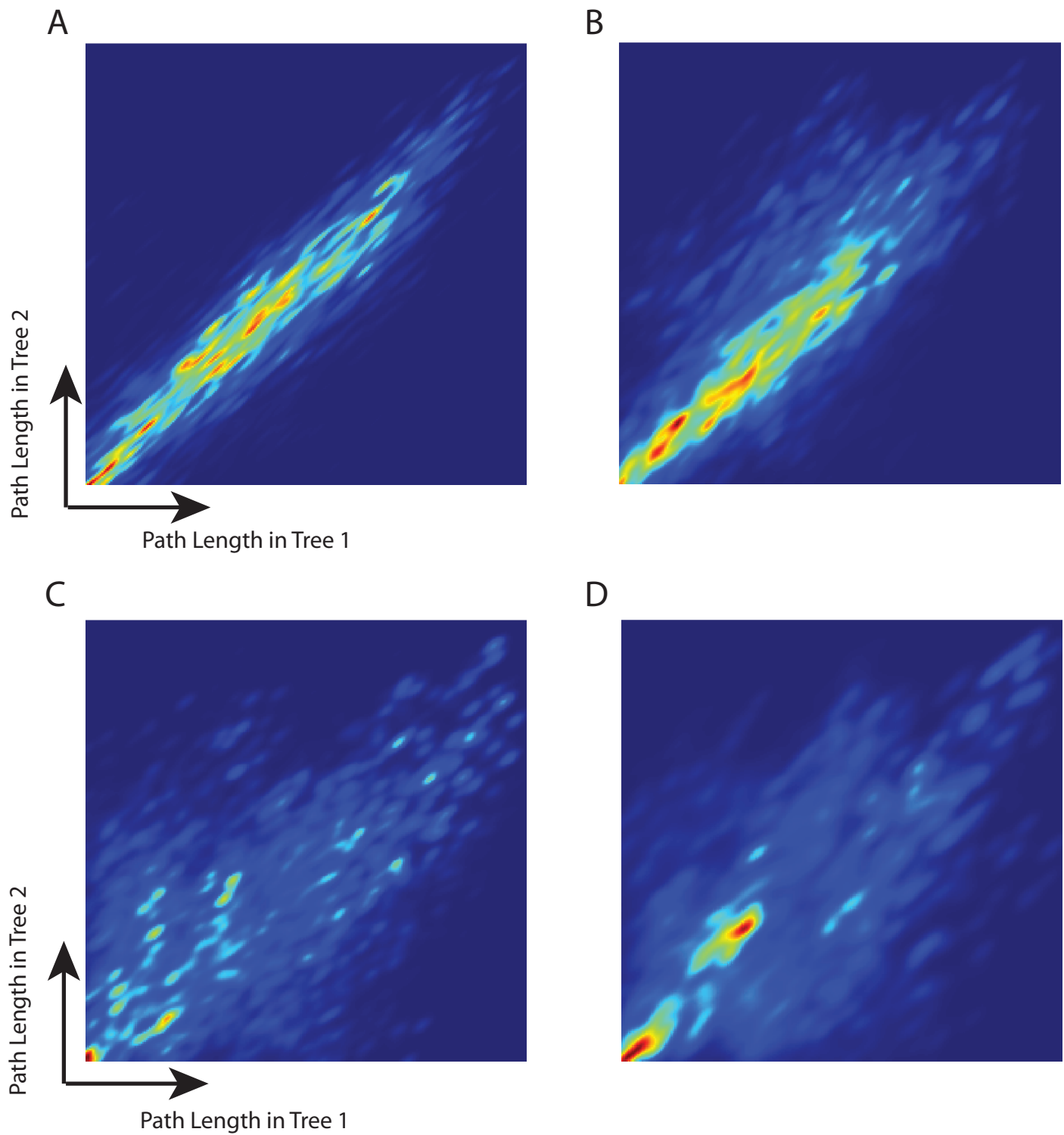

Figure 4

A
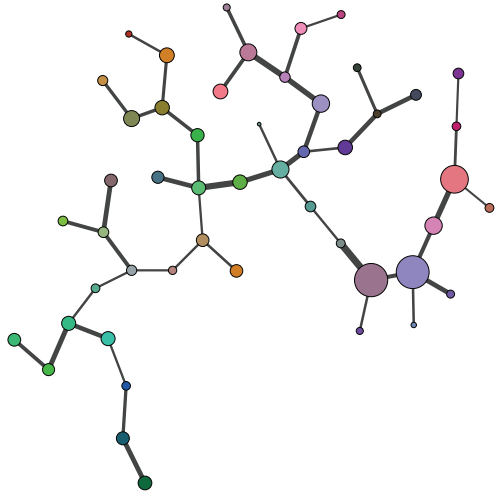

B
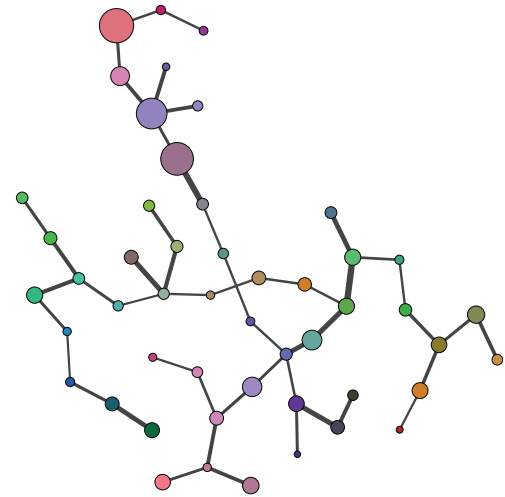

C
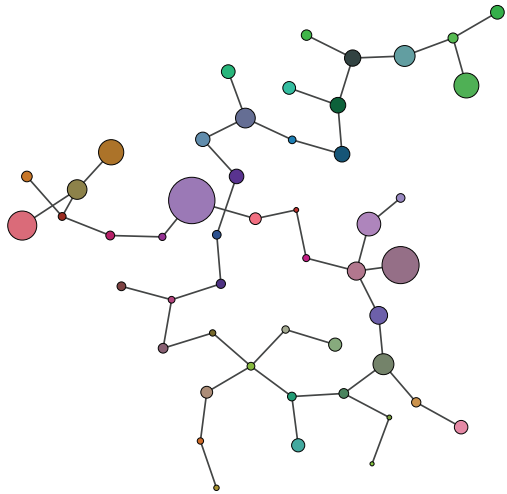

D
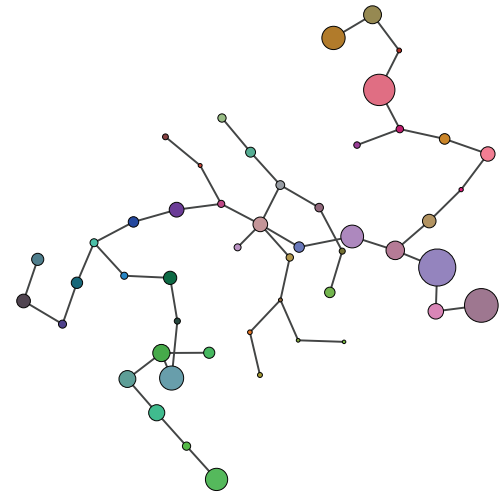

Figure 5

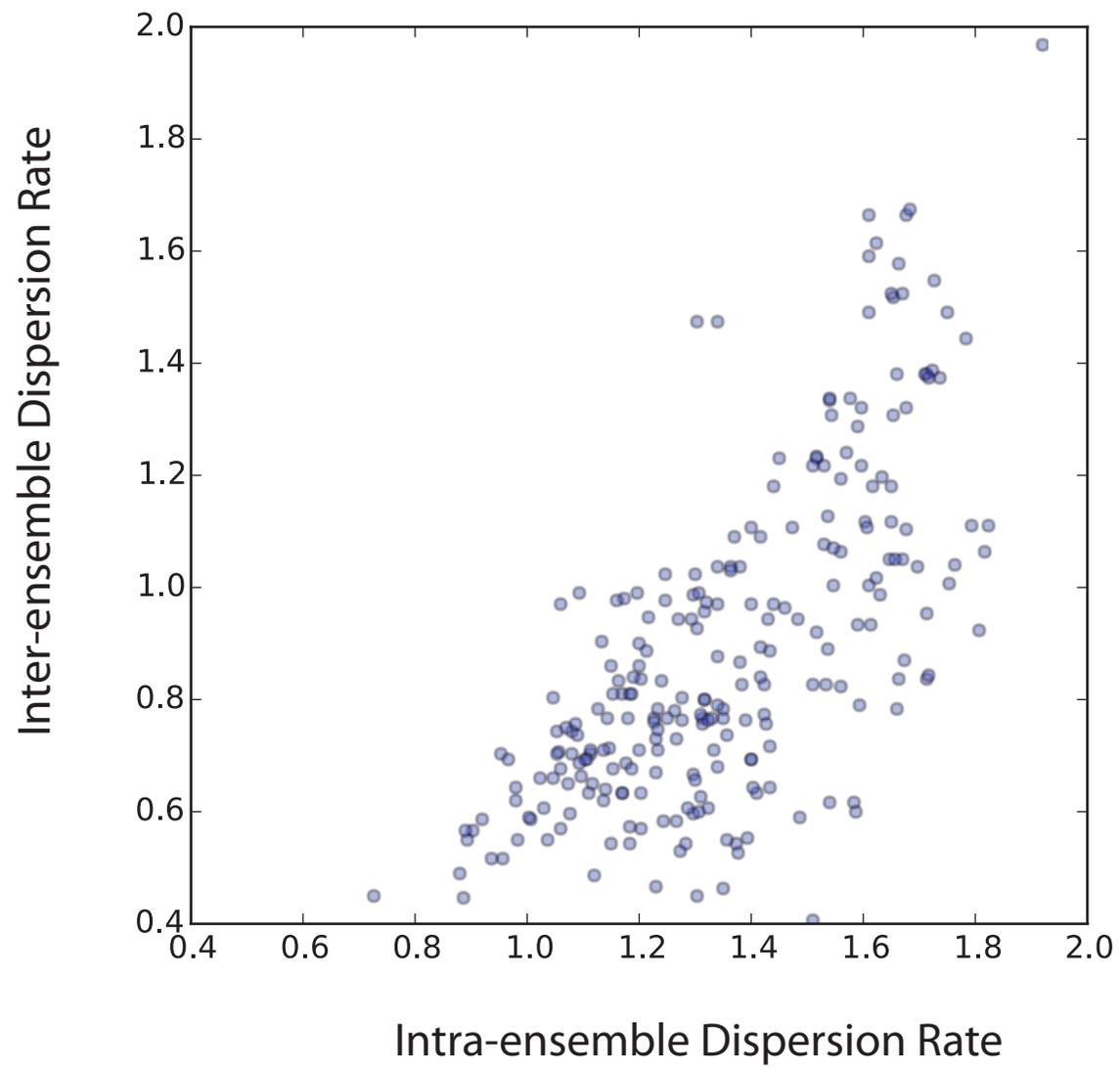

Figure S1

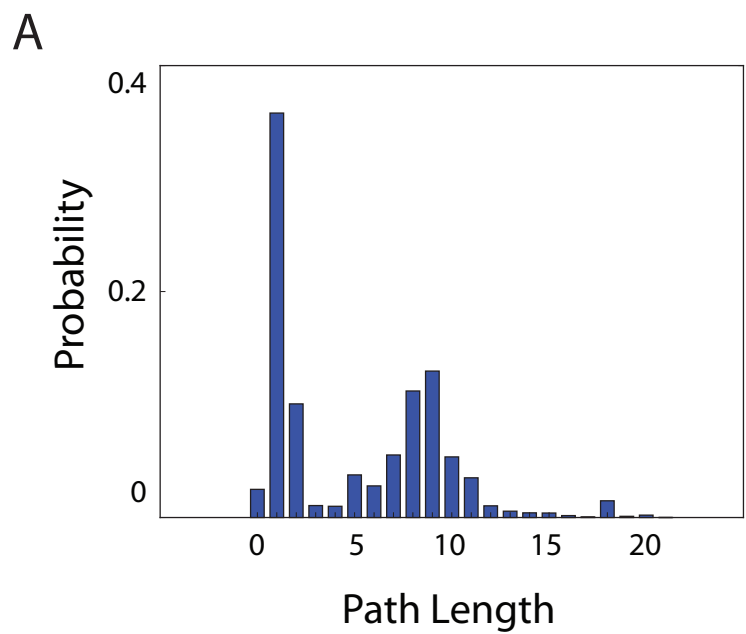
A

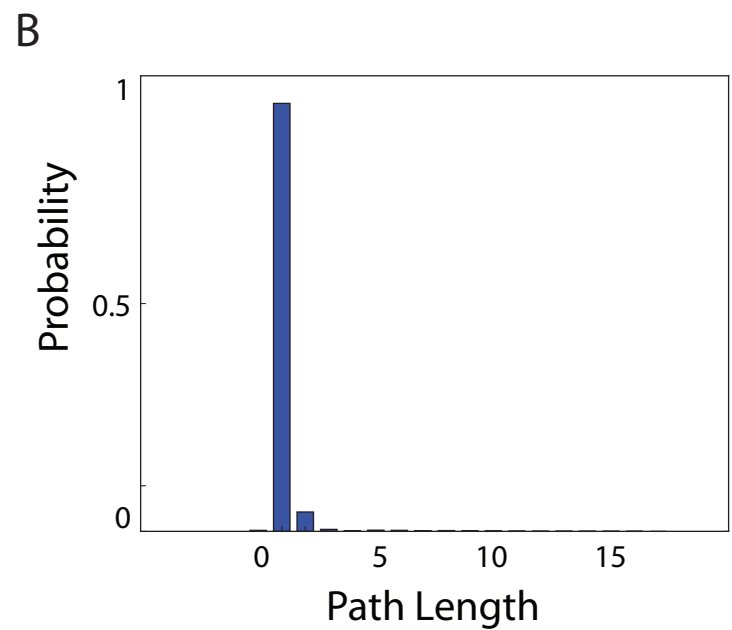
B

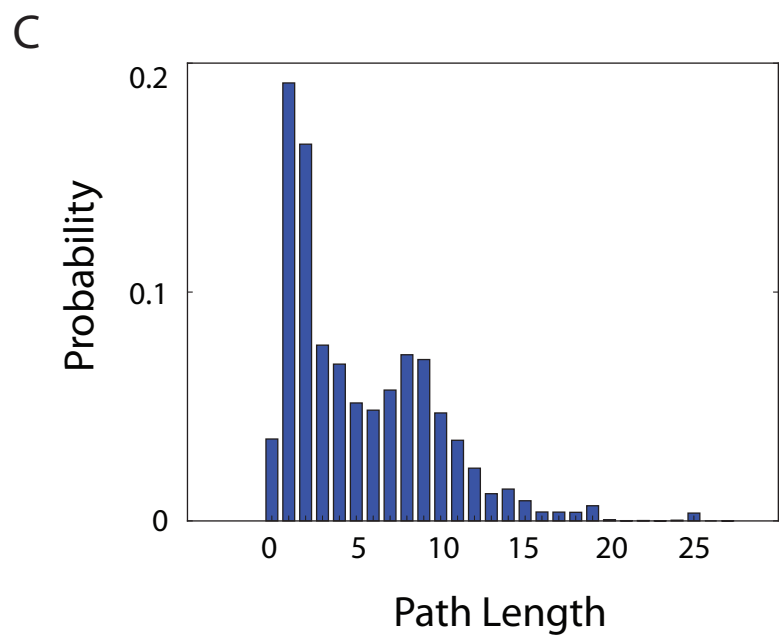
C

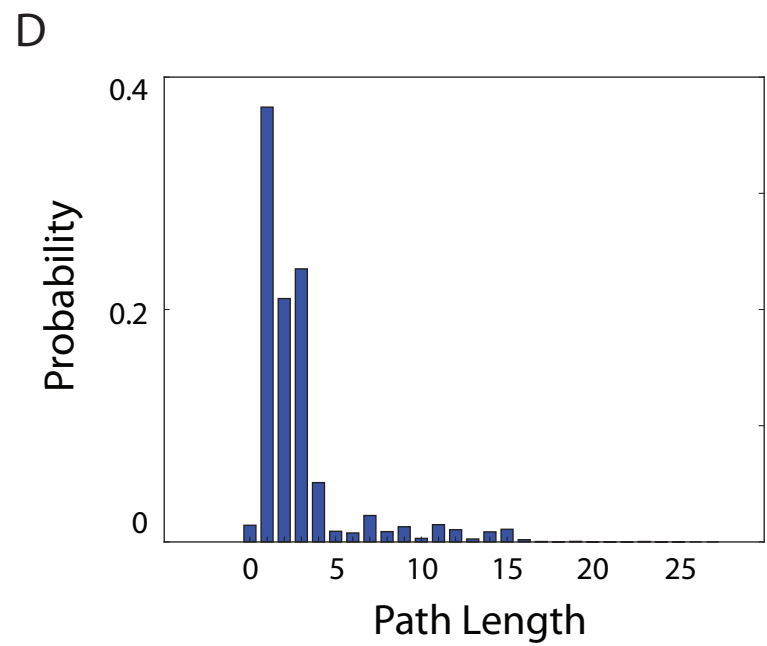
D

Figure S2

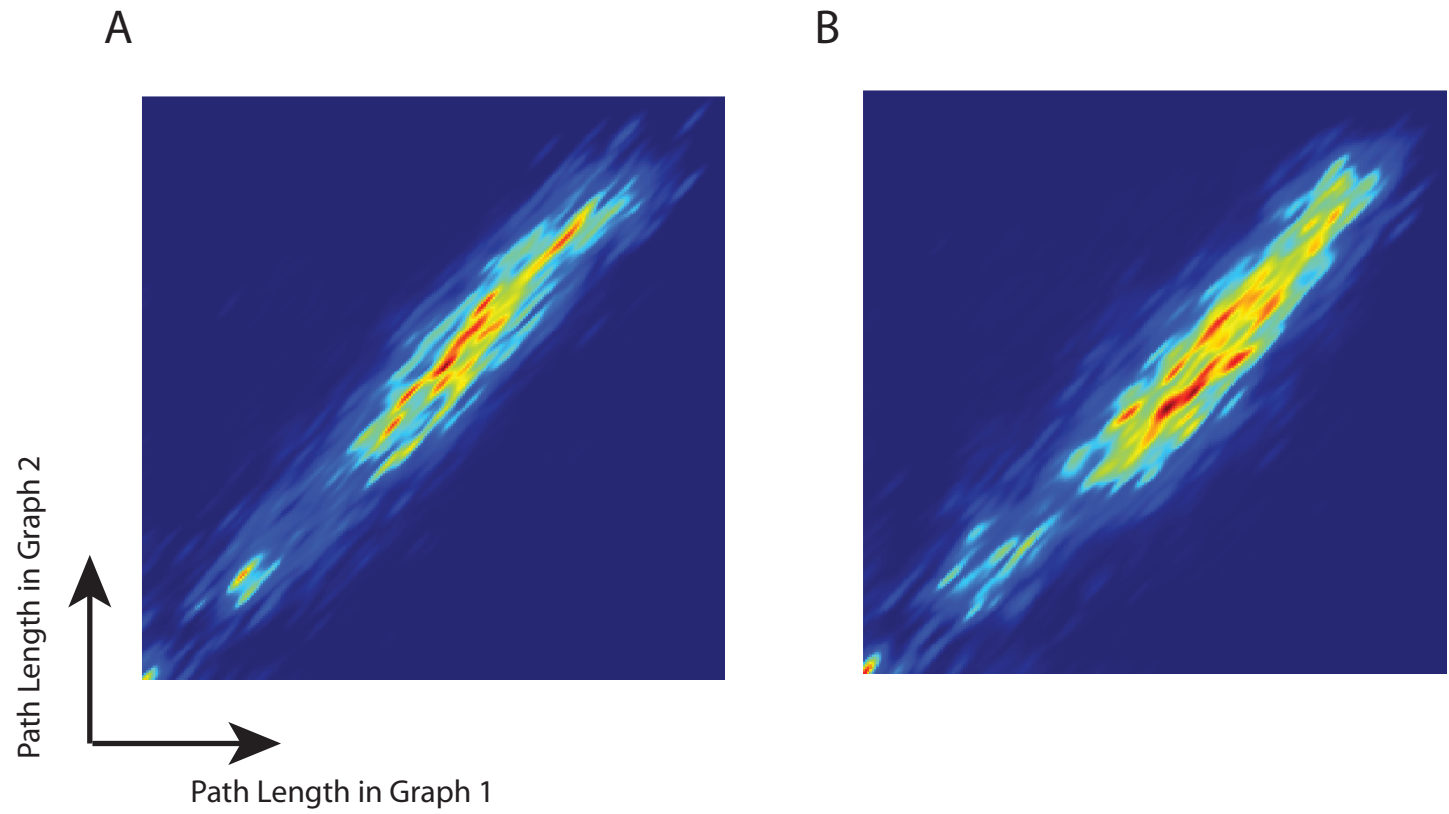

Figure S3

A 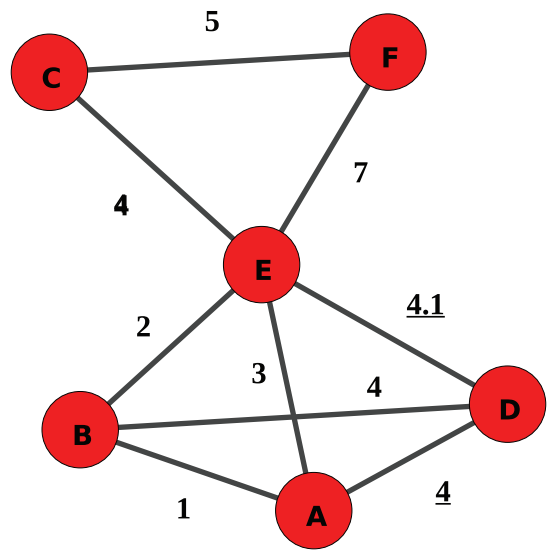

B 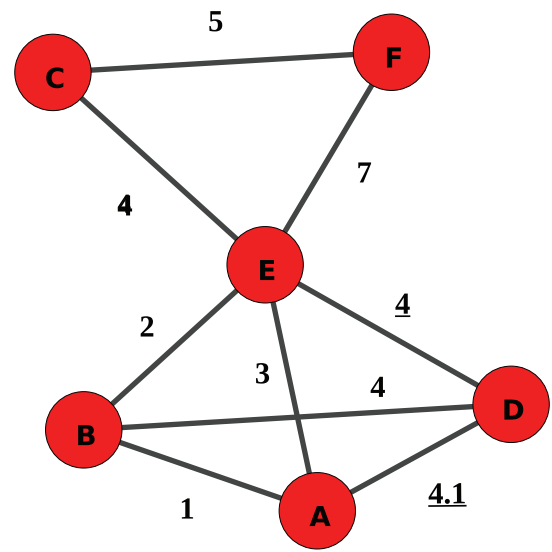

C 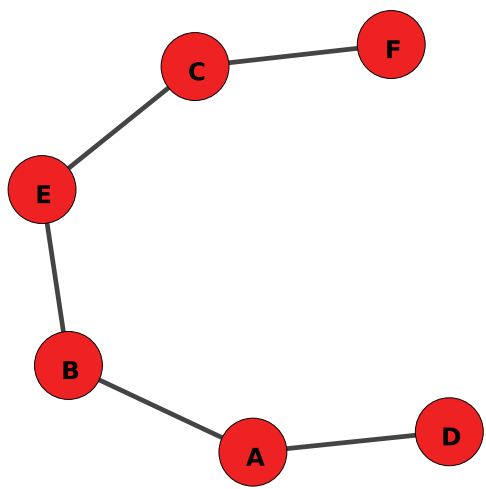

D 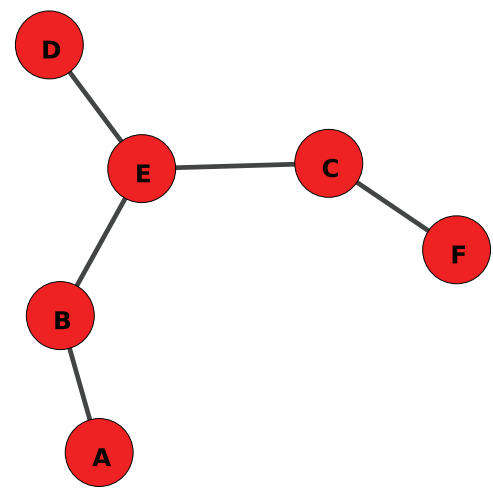

**FIGURE CAPTIONS**

**Figure 1: Overview of the ECLAIR method.** First, multiple subsamples are randomly drawn from the data. Each subsample is divided into cell clusters with similar gene expression patterns, and a minimum spanning tree is constructed to connect the cell clusters. Next, consensus clustering is constructed by aggregating information from all cell clusters. Finally, a lineage tree connecting the consensus clusters is constructed by aggregating information from the tree ensemble.

**Figure 2: ECLAIR correctly reconstructs the lineage tree in mouse early embryo.** (A). the lineage tree constructed by SCUBA, based on temporal information in the data. This tree has been experimentally validated. (B) The lineage tree constructed by ECLAIR, without using temporal information. The size of each node is proportional to the number of cells in the corresponding cluster. The color of each node indicates the gene expression pattern associated with the corresponding cell cluster. In (B), the edge width is inversely proportional to the estimated dispersion rate.

**Figure 3: Comparison of the reproducibility between ECLAIR (A-B) and SPADE (C-D).** Each heatmap shows the probability density of the cell-pair path length estimated using two trees obtained from the same method. (A). Two independent runs of ECLAIR on the same training set. (B). Two independent runs of ECLAIR on different training datasets. (C). Two independent runs of SPADE on the same training set. (D). Two Independent runs of SPADE on different training datasets.

**Figure 4: Comparison of the lineage tree structure between ECLAIR (A-B) and SPADE (C-D).** (A,B) Two independent runs of ÉCLAIR, the edge width is inversely proportional to the estimated dispersion rate. (C-D) Two independent runs of SPADE For (A) and (B),

**Figure 5: Correlation between the intra-ensemble and inter-ensemble dispersion rates.**

**Figure S1. Comparison between the distribution of intra-ensemble and inter-ensemble path lengths.** (A,B): Distribution of inter-ensemble path lengths for two randomly selected edges. (C,D): Distribution of intra-ensemble path lengths for the same edges.

**Figure S2: Reproducibility of the fully connected graph.** Each heatmap shows the probability density of the edge weight associated with two ECLAIR graphs. (A). Two independent runs of ECLAIR on the same training set. (B). Two independent runs of ECLAIR on different training datasets.

**Figure S3**: **A schematic showing the minimum spanning tree is sensitive to small perturbation of the edge weights.** (A,B) Two graphs with continuous edge weights. (C,D) The minimum spanning trees obtained from the graphs. differ significantly. Notice that while the two graphs have similar edge weight, the corresponding minimum spanning trees have different topologies.